\documentclass{article}
\usepackage{graphicx} 
\usepackage[utf8]{inputenc}
\usepackage{amsmath}
\usepackage{mathtools}
\usepackage{physics}
\usepackage{graphicx}
\usepackage{float}
\usepackage{bigints}
\usepackage{geometry}
\graphicspath{ {./images/} }

\usepackage{hyperref}
\hypersetup{
    colorlinks=true,
    linkcolor=blue,
    filecolor=magenta,      
    urlcolor=cyan,
    pdfpagemode=FullScreen,
}

\title{\Large \textbf{Newtonian and Post-Newtonian
Aspects of Mimetic Gravity}\vspace{25pt}}
\author{\textbf{Leonid Sarieddine} \vspace{15pt} \\ \small \textit{Physics department, American University of Beirut, Lebanon }  }

\date{}
\begin{document} 
\maketitle

\begin{abstract}
Mimetic gravity is a modified theory of gravity which is able to incorporate dark matter into the underlying geometry of space-time by isolating the conformal degree of freedom. The theory has been studied extensively in the cosmological regime, as such, we set out to study the implications of the theory at astrophysical scales. To that end, we carry out the post-Newtonian expansion of mimetic gravity to lowest post-Newtonian order. We interpret the equations in the Newtonian limit and study some of the implications of the theory at the solar system scale. Then by establishing some bounds on the asymptotic behavior of the fields we prove that any static spherically symmetric space-time with a non trivial mimetic contribution cannot be asymptotically flat. Finally, we study static spherically symmetric solutions. To explain the rotation curves, one needs a logarithmic term in the potential, we show that even though the mimetic fluid can't reconstruct an exact logarithmic term, it is able to contribute a quasi-logarithmic term which recovers the basic qualitative features of galactic rotation curves. 
\end{abstract}

\par\vspace{15pt}
\section{Introduction}
Despite its immense success, it’s well known that general relativity is not an entirely satisfactory theory and faces several problems. To that end, many alternate theories have been proposed throughout the decades, one of the most recent ones being the so called mimetic gravity which was proposed by Chamseddine and Mukhanov \cite{chamseddine2013mimetic}. The starting point of the theory is a parametrization of the physical metric by an auxiliary metric and a scalar field in such a way that the physical metric remains invariant under conformal transformations of the auxiliary metric isolating the conformal mode of gravity in a covariant fashion. The resulting equations of motion lead to more general solutions than those in standard GR and are able to incorporate dark matter into the gravitational metric degrees of freedom in a very natural way. Similar theories were also considered before then (see in particular \cite{lim2010dust}).

Later on, it was shown that an equivalent (and more convenient) formulation could be established by a Lagrange multiplier approach, essentially by adding a Lagrange multiplier which constrains the gradient of the mimetic field to be timelike \cite{golovnev2014recently}. It then becomes easy to see that mimetic gravity is a form of scalar Einstein aether theory \cite{sebastiani2017mimetic}. By modifying the mimetic gravity action to include a potential which depends solely on the mimetic field one can show that it can mimic inflationary periods, quintessence, bouncing universe, and essentially any background cosmology that one desires \cite{chamseddine2014cosmology}. By introducing further terms in the action one can also resolve cosmological singularities \cite{chamseddine2017resolving} as well as black hole singularities \cite{chamseddine2017nonsingular}. 

This work is concerned with studying the implications of mimetic gravity (without any additional terms in the action) on the solar system and galactic scales, more precisely in the context of post-Newtonian expansion. The post-Newtonian expansion is well known in standard general relativity and it is responsible for calculating the effects of the theory in the weak field and weak velocity limit where gravity is weak and the motion of the bodies are slow enough such that systematic expansions in inverse powers of characteristic length scales and inverse powers of speed of light can be carried out. 

The results are organized as follows: In section 2, we carry out the post-Newtonian expansion in mimetic gravity obtaining the governing equations to lowest post-Newtonian order. We proceed to section 3 by noting that in the Newtonian limit mimetic gravity is reduced to a system of equations akin to Newton’s gravity coupled to hydrodynamic pressure-less ideal Eulerian fluid whose velocity profile comes from the gradient of the mimetic field. However, the analogy is not exact since the fluid represents an intrinsic gravitational degree of freedom as opposed to being a source of gravity. In section 4 we explore some of the implications of the theory where we point out that unlike on the cosmological scale, the mimetic fluid can’t mimic the exact dark matter profile on astrophysical scales. In section 5 we prove that any static spherically symmetric space-time with a nontrivial mimetic contribution cannot be asymptotically flat, essentially due to the contribution of the mimetic energy density. In section 6 we look for explicit solutions in the Newtonian limit of static spherically symmetric spacetimes using a perturbative as well as numerical approaches which are able to reconstruct the basic qualitative features of rotation curves. 

\section{Deriving the Equations}

The post-Newtonian expansion of the metric tensor and by extension the Ricci, Riemann and Einstein tensors are standard and well known (see \cite{straumann2012general} or \cite{weinberg1972gravitation}). We follow the conventions of the former. The equations of mimetic gravity that are to be expanded are:

$$G_{\mu \nu} +(G-T)\partial_{\mu}\phi \partial_{\nu}\phi=T_{\mu \nu}$$

$$\partial_{\mu}(\sqrt{-g}(G-T)g^{\mu \nu} \partial_{\nu}\phi)=0$$

$$g^{\mu \nu}\partial_{\mu}\phi \partial_{\nu}\phi=-1$$ Obviously we expect that the $G-T$ term to be expanded as:

$$G-T= {}\stackrel{(2)}{G-T}+ \stackrel{(4)}{G-T}+ \stackrel{(6)}{G-T}+ ....$$ where $\stackrel{(n)}{G-T}$ is of the order of $\dfrac{v^n}{R^2}$.\par\vspace{5pt}\noindent As for $\phi$, it is to be expanded in the following manner:

$$\phi = t + \stackrel{(1)}{\phi}+ \stackrel{(3)}{\phi}+ \stackrel{(5)}{\phi}+...$$where $\stackrel{(n)}{\phi}$ is of the order of $Rv^n$. Perhaps a more suggestive way to look at the above expansion is to note that $\phi$ only ever appears in terms of derivatives and hence it's more suggestive to write the expansion of $\phi$ in this case:

$$\partial_{i}\phi={} \stackrel{(1)}{\partial_{i}\phi}+\stackrel{(3)}{\partial_{i}\phi}+\stackrel{(5)}{\partial_{i}\phi}+...$$where now $\stackrel{(n)}{\partial_{i}\phi}$ is of order $v^n$, and

$$\partial_{0}\phi= 1+\stackrel{(2)}{\partial_{0}\phi}+\stackrel{(4)}{\partial_{0}\phi}+\stackrel{(6)}{\partial_{0}\phi}$$ where $\stackrel{(n)}{\partial_{0}\phi}$ is of order $v^n$.
\par\vspace{5pt}\noindent Now we begin by expanding the constraint equation. Using the expansion of the inverse metric and $\phi$ we compute:

$$g^{00}\partial_{0}\phi\partial_{0}\phi+2g^{0i}\partial_{0}\phi\partial_{i}\phi+g^{ij}\partial_{i}\phi\partial_{j}\phi=-1$$
\begin{gather*}
  (-1+\stackrel{(2)}{g^{00}}+\stackrel{(4)}{g^{00}}+...)(1+\stackrel{(2)}{\partial_{0}\phi}+\stackrel{(4)}{\partial_{0}\phi}+...)(1+\stackrel{(2)}{\partial_{0}\phi}+\stackrel{(4)}{\partial_{0}\phi}+...) \\
   +2(\stackrel{(3)}{g_{0i}}+\stackrel{(5)}{g_{0i}}+...)(1+\stackrel{(2)}{\partial_{0}\phi}+\stackrel{(4)}{\partial_{0}\phi}+...)(\stackrel{(1)}{\partial_{i}\phi}+\stackrel{(3)}{\partial_{i}\phi}+\stackrel{(5)}{\partial_{i}\phi}+...)\\
+(\delta^{ij}+\stackrel{(2)}{g^{ij}}+\stackrel{(4)}{g^{ij}})(\stackrel{(1)}{\partial_{i}\phi}+\stackrel{(3)}{\partial_{i}\phi}+\stackrel{(5)}{\partial_{i}\phi}+...)(\stackrel{(1)}{\partial_{j}\phi}+\stackrel{(3)}{\partial_{j}\phi}+\stackrel{(5)}{\partial_{j}\phi}+...)=-1
\end{gather*}We will expand the above equation to second order only, hence we obtain:

$$-1+\stackrel{(2)}{g^{00}}-2 \stackrel{(2)}{\partial_{0}\phi}+\delta^{ij}\stackrel{(1)}{\partial_{i}\phi}\stackrel{(1)}{\partial_{i}\phi}{}=-1$$Hence the final equation which needs to be satisfied at second order is:

$$-\dfrac{1}{2}\stackrel{(2)}{g_{00}}+\dfrac{1}{2}|\stackrel{(1)}{\nabla{\phi}}|^2 ={} \stackrel{(2)}{\partial_{0}\phi}$$Where we have used the fact that $\stackrel{(2)}{g_{00}}{}=-\stackrel{(2)}{g^{00}}$\par\vspace{15pt}\noindent Now we expand the second equation as follows:

$$\partial_{0}(\sqrt{-g}(G-T)g^{0 \nu} \partial_{\nu}\phi)+\partial_{i}(\sqrt{-g}(G-T)g^{i \nu} \partial_{\nu}\phi)=0$$Note that the determinant can be expanded as:

$$-g=1+\stackrel{(2)}{g}+ ...$$where $\stackrel{(2)}{g}$ is of order $v^2$ and so on. Hence the square root can be expanded as:

$$\sqrt{-g}=\sqrt{1+\stackrel{(2)}{g}+...}=1+\dfrac{1}{2}\stackrel{(2)}{g}+...$$Meanwhile the term $g^{0 \nu} \partial_{\nu}\phi$ can be expanded as:

\begin{gather*}
g^{0 \nu} \partial_{\nu}\phi= \bigg(-1 +\stackrel{(2)}{g^{00}}+\stackrel{(4)}{g^{00}}+...\bigg)\bigg(1+\stackrel{(2)}{\partial_{0}\phi}+\stackrel{(4)}{\partial_{0}\phi}+...\bigg)+\bigg(\stackrel{(3)}{g^{0i}}\\+\stackrel{(5)}{g^{0i}}+...\bigg)\bigg(\stackrel{(1)}{\partial_{i}\phi}+\stackrel{(3)}{\partial_{i}\phi}+\stackrel{(5)}{\partial_{i}\phi}+...\bigg)
\end{gather*} We only expand this to order 2:

$$g^{0 \nu} \partial_{\nu}\phi= -1+\stackrel{(2)}{g^{00}}+ \stackrel{(2)}{\partial_{0}\phi}+... $$\\ Meanwhile as for the $g^{i \nu} \partial_{\nu}\phi$ term:

\begin{gather*}g^{i \nu} \partial_{\nu}\phi= \bigg(\stackrel{(3)}{g^{0i}}+\stackrel{(5)}{g^{0i}}+...\bigg)\bigg(1+\stackrel{(2)}{\partial_{0}\phi}+\stackrel{(4)}{\partial_{0}\phi}+...\bigg)+\bigg(\delta^{ij}+\\\stackrel{(2)}{g^{ij}}+\stackrel{(4)}{g^{ij}}+...\bigg)\bigg(\stackrel{(1)}{\partial_{j}\phi}+\stackrel{(3)}{\partial_{j}\phi}+\stackrel{(5)}{\partial_{j}\phi}+...\bigg)\end{gather*}Finally we obtain:

$$g^{i \nu} \partial_{\nu}\phi={} \stackrel{(1)}{\partial_{i}\phi}+(\stackrel{(3)}{\partial_{i}\phi}+\stackrel{(2)}{g^{ij}}\stackrel{(1)}{\partial_{j}\phi}+\stackrel{(3)}{g^{i0}})+...$$ Hence the full expansion is:

\begin{gather*}
 \partial_{0}\Bigg(\bigg(1+\dfrac{1}{2}\stackrel{(2)}{g}+...\bigg)\bigg(\stackrel{(2)}{G-T}+....\bigg)\bigg(-1+\stackrel{(2)}{g^{00}}+ \stackrel{(2)}{\partial_{0}\phi}+...\bigg)\Bigg)\\
 +\partial_{i}\Bigg(\bigg(1+\dfrac{1}{2}\stackrel{(2)}{g}+...\bigg)\bigg(\stackrel{(2)}{G-T}+....\bigg)\bigg(\stackrel{(1)}{\partial_{i}\phi}+(\stackrel{(3)}{\partial_{i}\phi}+\stackrel{(2)}{g^{ij}}\stackrel{(1)}{\partial_{j}\phi}+\stackrel{(3)}{g^{i0}})+...\bigg)\Bigg)=0
\end{gather*}To extract the lowest order term recall that spatial derivatives are weighed differently than time derivatives. The term $\partial_{0}\big(\stackrel{(2)}{G-T}\big)$ is clearly of order $\dfrac{v^3}{R^3}$. The product of $\stackrel{(2)}{(G-T)}\stackrel{(1)}{\partial_{i}\phi}$ is of order $\dfrac{v^3}{R^2}$ and a spatial derivative introduces an extra factor of $1/R$, hence these two terms are of the same order; so to lowest order this equation reads:

$$\partial_{0}\stackrel{(2)}{(G-T)}{}
 =\partial_{i}\big(\stackrel{(2)}{(G-T)}\stackrel{(1)}{\partial_{i}\phi}\big)$$ This almost looks like a continuity equation and in fact will be whenever we interpret $\phi$ properly later on.
\par\vspace{15pt}
 \noindent Now we expand the modified Einstein equation. To do that we first expand the $(G-T)\partial_{\mu}\phi \partial_{\nu}\phi$ term:

 $$(G-T)\partial_{0}\phi \partial_{0}\phi= (\stackrel{(2)}{G-T}+ \stackrel{(4)}{G-T})(1+\stackrel{(2)}{\partial_{0}\phi}+\stackrel{(4)}{\partial_{0}\phi})(1+\stackrel{(2)}{\partial_{0}\phi}+\stackrel{(4)}{\partial_{0}\phi})$$ hence it's easily seen that to lowest order

$$(G-T)\partial_{0}\phi \partial_{0}\phi= ( \stackrel{(2)}{G-T})+...$$ Similarly the other terms to lowest order are:

$$(G-T)\partial_{0}\phi \partial_{i}\phi= ( \stackrel{(2)}{G-T})\stackrel{(1)}{\partial_{i}\phi}+...$$

$$(G-T)\partial_{i}\phi \partial_{j}\phi= ( \stackrel{(2)}{G-T})\stackrel{(1)}{\partial_{i}\phi}\stackrel{(1)}{\partial_{j}\phi}+...$$Note that the space-space term starts at order 4 and hence won't contribute to lowest order field equations. Hence finally we can write down the field equations to lowest order; plugging in the expansion of the Einstein tensor previously derived we get the following:

$$\dfrac{1}{2}\stackrel{(2)}{R_{00}}+\dfrac{1}{2}\stackrel{(2)}{R_{jj}}+( \stackrel{(2)}{G-T})={}\stackrel{(2)}{T_{00}}$$

$$\stackrel{(3)}{R_{0i}}+( \stackrel{(2)}{G-T})\stackrel{(1)}{\partial_{i}\phi}{}={}\stackrel{(3)}{T_{0i}}$$

$$\stackrel{(2)}{R_{ij}}+\dfrac{1}{2}\big(\stackrel{(2)}{R_{00}}-\stackrel{(2)}{R_{kk}}\big)\delta_{ij}={}\stackrel{(2)}{T_{ij}}$$\\If we write these explicitly in terms of the metric  (and note that $\stackrel{(2)}{T_{ij}}{}=0$ from standard post-Newtonian theory) then:

$$-\dfrac{1}{4}\Delta{\stackrel{(2)}{g_{00}}}-\dfrac{1}{4}\Delta{\stackrel{(2)}{g_{jj}}}={}\stackrel{(2)}{T_{00}}-( \stackrel{(2)}{G-T})$$

$$-\dfrac{1}{2}\Delta{\stackrel{(3)}{g_{0i}}}={}\stackrel{(3)}{T_{0i}}-{(G-T)}\stackrel{(1)}{\partial_{i}\phi}$$

$$-\dfrac{1}{2}\Delta{\stackrel{(2)}{g_{ij}}}+\dfrac{1}{4}(-\Delta{\stackrel{(2)}{g_{00}}}+\Delta{\stackrel{(2)}{g_{kk}}})\delta_{ij}=0$$It is possible to further simplify the equations; taking the 3-trace of the last equation we obtain:

$$\dfrac{3}{4}\Delta{\stackrel{(2)}{g_{00}}}=\dfrac{1}{4}\Delta{\stackrel{(2)}{g_{kk}}}$$plugging this back into the first equation we obtain:

$$\Delta{\stackrel{(2)}{g_{00}}}=-\stackrel{(2)}{T_{00}}+( \stackrel{(2)}{G-T})$$Finally, we arrive at the Post-Newtonian Mimetic equations as:

$$\Delta{\stackrel{(2)}{g_{00}}}=-\stackrel{(2)}{T_{00}}+( \stackrel{(2)}{G-T})$$

$$-\dfrac{1}{2}\Delta{\stackrel{(3)}{g_{0i}}}={}\stackrel{(3)}{T_{0i}}-{(G-T)}\stackrel{(1)}{\partial_{i}\phi}$$

$$\partial_{0}\stackrel{(2)}{(G-T)}
 {}=\partial_{i}\big(\stackrel{(2)}{(G-T)}\stackrel{(1)}{\partial_{i}\phi}\big)$$

$$-\dfrac{1}{2}\stackrel{(2)}{g_{00}}+\dfrac{1}{2}|\stackrel{(1)}{\nabla{\phi}}|^2 = {}\stackrel{(2)}{\partial_{0}\phi}$$

\section{Interpreting The Equations}

We will be working with the 4 equations above and hence to simplify notation we define $g_{00}:=-2U$ $E := \frac{(G-T)}{2}$, and $\psi:=-\phi$, all evaluated to lowest order.

\noindent Notice that if we take the gradient of the last equation we obtain:

$$ \nabla{\nabla{\psi}}\cdot\nabla{\psi} +{\partial_{0}\nabla\psi}= -\nabla{U}$$Where $\nabla\nabla\psi$ is the Hessian of $\psi$ and $\cdot$ is the vector tensor dot product. Notice that this is exactly the Euler equation for a presureless fluid with $\nabla\psi$ playing the role of a velocity field. The left hand side represents the convective derivative of the velocity field and the RHS is the gravitational force. Of course this could have also been derived by the conservation of energy momentum tensor (and expanding it to lowest order). We also have a continuity equation:\\

$$\partial_{0}E
 =-\nabla \cdot (E\nabla\psi)$$ We see the somewhat expected final result that mimetic gravity is equivalent to Newtonian gravity coupled to an ideal pressure-less fluid whose velocity field comes from a potential function. This fluid, however, is of course not part of any regular matter but may be thought of as an intrinsic component of gravity. 

\section{Some Interesting Implications}

Let us re-write the 3 main equations (we drop the space-time Einstein equation since it plays no role in what follows) where we also reinstate relevant constant:

\begin{equation}\label{eq:test1} \Delta U=4\pi G \rho - E\end{equation}

\begin{equation}\label{eq:test2}\partial_{0}E
 =-\nabla \cdot (E\nabla\psi)\end{equation}

\begin{equation}\label{eq:test3}\frac{U}{c^2}+\dfrac{1}{2}\|\nabla{\psi}\|^2 = - \partial_{0}\psi\end{equation}Notice that this is a system of 3 non-linear PDE's for 3 unknowns provided that $\rho$ is given; if not then we would couple it to the standard hydrodynamic matter. So even if the matter density is given, one can't in general write a closed form solution for the potential $U$ since it is also determined by $E$ which in turn is coupled to $\psi$ which in turn is coupled back to the potential $U$. We notice that if $E=0$ then \eqref{eq:test2} is identically satisfied and the solution for $U$ agrees with the standard GR/Newtonian solution. Then $\psi$ is determined in terms of $U$ via \eqref{eq:test3}. In this case, $\psi$ doesn't directly affect $U$ neither the dynamics of particles in the Newtonian limit (which are only sensitive to $U$; this can be seen from the expansion of geodesic equation). Also the fact that $E$ is subtracted from $\rho$ is not particularly significant since in practice its sign is determined by an integration constant.\\

\noindent To obtain solutions not equivalent to GR we must have $E\ne 0$. Let us begin by assuming that $E$ is time independent, then \eqref{eq:test2} implies that a general class of solutions for $E\nabla \psi$ is given by:

\begin{equation}\label{eq:test4}E \nabla \psi= \nabla \times \vec{A}\end{equation}$\vec{A}$ is a free parameter of the theory and may be thought as representing the amount of mimetic matter as well as its direction (in the same spirit as when it was originally conceived to mimic cold dark matter in the original paper). Hence $\vec{A}$ immediately implies a preferred direction in space or a preferred frame. This is of course not surprising because there is a well known link between Mimetic gravity and Einstein aether theories as mentioned in the introduction. We will eventually be interested in applying the theory to explain the effects of dark matter on astrophysical scales. \eqref{eq:test4} also allows us to immediately write down the solution to $g_{0i}$:

$${\stackrel{(3)}{g_{0i}}}=-2 \int \stackrel{(3)}{T_{0i}}-(\nabla \times \vec{A})_i $$Note that mimetic gravity's original success was attributed to it being able to incorporate cosmological dark matter into the geometry of space-time, so the natural question is whether it is able to do so on astrophysical scales. After all, the original evidence for dark matter came from analyzing the so called rotation curves of galaxies and it was realized that the visible matter couldn't account for the observations \cite{arbey2021dark}, hence the standard solution to the problem is to postulate the existence of an extra component of matter called dark matter (the other being to modify the gravitational theory itself, as for example is done in MOND). Mimetic gravity is able to incorporate dark matter into the geometry of space-time on cosmological scales where it is assumed that space-time (and hence, by extension, matter) is homogeneous and isotropic. This is a very special setting and possesses a very high degree of symmetry; when we go down to astrophysical scales (where Newtonian limit still applies), we observe that mimetic gravity is unable to incorporate the effects of dark matter into the mimetic fluid. To see this we first note that to obtain the required rotation curve on galactic scale it is required that the density profile of $E$ to satisfy $$E \propto \frac{1}{r^2}$$ outside the regular matter (where $\rho = 0$). From the continuity equation we see that $$E\psi' \propto \frac{1}{r^2}$$ Hence we see that to get the required profile we must have $\psi' = \text{constant}$. But \eqref{eq:test3} then implies that $U$ is also a constant (assuming a static case). On the other hand, using \eqref{eq:test1} we get
$$U=-GM/r + \text{term linear in $r$}$$ Clearly, contradicting the existence of such a profile. So we see that on astrophysical scales mimetic gravity can't mimic the exact profile dark matter.\\

A natural question at this point would be the following: What kind of constraints can we put on the mimetic matter/fluid such that it agrees with solar system tests? The general method to answer such a question would be to compute the PPN parameters of the theory and relate them to the experimentally known values. Unfortunately, mimetic gravity doesn't seem to fit in the parametrized post-Newtonian framework as established, say, in \cite{will2018theory}, hence more direct methods are required. We can, however, say something about a static space-time.
As mentioned in the introduction, previous work on static space-times in mimetic gravity assumed that the mimetic field is independent of time. In our work, we expanded $\phi$ around the solution $\phi =t$, hence it is not time-independent, but this is of no harm since one notices that only gradients of $\phi$ enter the equations, hence we can still study static space-times. \\

We find that any static solution would necessarily lead to a different Newtonian limit than standard GR. To see this, a static space-time implies the solution to \eqref{eq:test2} is given by \eqref{eq:test4} which implies that:

$$E^2\|\nabla{\psi}\|^2=\|\nabla \times \vec{A}\|^2$$Then \eqref{eq:test3} implies that (assuming $U<0$ which it is in regular Newtonian case).

\begin{equation}\label{eq:test5}   \|\nabla{\psi}\|^2 = -\frac{2U}{c^2}\end{equation}Then assuming $E>0$ (energy density of mimetic fluid is positive) we have 

$$E=\dfrac{c\|\nabla \times \vec{A}\|}{\sqrt{-2U}}$$Plugging this back in \eqref{eq:test1} we obtain:

$$\Delta U= 4\pi G \rho - \dfrac{c\|\nabla \times \vec{A}\|}{\sqrt{-2U}} $$ Clearly this will lead to a solution quite different than the standard Newtonian case, and the deviation could be thought of as being measured by the free parameter $\vec{A}$ (if $\vec{A}$ or $\nabla \times \vec{A} =0$ we recover GR). Hence we conclude that all non-GR static solutions (with $E \ne 0$) necessarily lead to deviations from the standard Newtonian potential. This is to be contrasted to the Einstein vector theory where one can obtain a solution in which the aether field is in the direction of the killing vector $\partial_{t}$ and the result would just be a modification of the effective gravitational constant \cite{carroll2004lorentz}. But the proportionality constant between geometry and matter is determined by the Newtonian limit hence there's no contradiction with Newtonian theory. Such a thing is not possible in mimetic gravity because, among other things, a mimetic fluid being in the direction of the time-like Killing vector contradicts \eqref{eq:test3}.

\section{Lack of Asymptotic Flatness}

Here we shall show that any static spherically space-time cannot be asymptotically flat . Let us take our domain to be $D=\{x \in R^3 : \norm{x} \ge R\}$ for some radius $R$. We will prove the claim by a contradiction. Suppose that space-time is asymptotically flat; this implies $U \to 0$ as $r \to \infty$; then \eqref{eq:test5} implies $\norm{\nabla \psi} \to 0$ as well. So there exist a constant $b$ and a radius $R'$ such that $\dfrac{1}{\norm{\nabla \psi}} \ge b$ for all $r \ge R'$\footnote{Without loss of generality we assume that the matter density vanishes exterior to $R'$}.

\noindent Static space-time implies $E$ is time-independent and hence

$$\nabla \cdot (E\nabla\psi) = 0$$Using the divergence theorem, this implies that for any spherical surface of radius $r$ (denoted by $S_{r})$ we have:

$$\oint_{S_{r}} E\nabla\psi \cdot \,ds = \oint_{S_{R}} E\nabla\psi \cdot \,ds = L$$where $L$ is a constant. If $r \ge R'$ and if $E$ doesn't change sign (which is reasonable since $E$ is an energy density analogue) then we have

$$|L|=\bigg| \oint_{S_{r}} E\nabla\psi \cdot \,ds \bigg|= \big| 4\pi r^2 \psi' E \big|$$ Hence we have the following bound, for $r \ge R'$:

$$ b|L| \le  \big|4\pi r^2 E \big|$$Now we turn to $U$. By using Gauss law we have

$$U'(r)4\pi r^2=  \bigintssss_{\norm{x'}\le R'} (\rho-E) \,d^3x' - \bigintssss_{r \ge \norm{x'}\ge R'} E \,d^3x'$$

$$U'(r)=\dfrac{1}{4 \pi r^2} \bigg(\bigintssss_{\norm{x'}\le R'} (\rho-E) \,d^3x' - \bigintssss_{r \ge \norm{x'}\ge R'} E \,d^3x' \bigg)$$The first integral clearly leads to a contribution to $U$ (not $U'$) that vanishes at $\infty$ hence we concentrate on the second one:

$$\dfrac{1}{4 \pi r^2} \bigg| \bigintssss_{r \ge \norm{x'}\ge R'} E \,d^3x' \bigg| = \dfrac{1}{4 \pi r^2} \bigg| \bigintssss_{R'}^{r} E 4 \pi r'^2 \,dr' \bigg|   \ge  \dfrac{b|L|}{4 \pi r^2} (r-R')= \dfrac{b|L|}{4 \pi r} - \dfrac{b|L|R'}{4 \pi r^2}$$Hence by integrating the above expression we see that as $r \to \infty$ we have:

$$|U(r)| \ge \dfrac{b|L|}{4\pi} \ln(r) \pm O(\dfrac{1}{r})$$leading to a contradiction. Essentially this says that the right hand side term $E$ in \eqref{eq:test1} doesn't decay fast enough to have an asymptotically flat space-time. Or in pure Newtonian terms, the mimetic energy density $E$ doesn't decay fast enough to have a vanishing potential at infinity.\\

\noindent This would imply that mimetic (static) black holes can't be asymptotically flat, since if they were, then in the asymptotic region the post Newtonian expansion would be valid. The argument above, however, shows that in the asymptotic region the contribution of the mimetic fluid is always significant enough to guarantee a non-asymptotically flat solution, which leads to a contradiction. This is in agreement with \cite{sebastiani2017mimetic} \cite{gorji2020mimetic} (although it should be noted that neither is working within the exact same framework as we are).

\section{Static Spherically Symmetric Case: Solutions}

Let us now study the physically most interesting situation where we assume static spherically symmetric space-time. As pointed out before, spherically symmetric space-times were already studied, but only with space-like constraints or complex scalar field. We study real solutions and seek to qualitatively explain rotation curves. Even though we already showed that an exact $1/r^2$ profile can't be obtained, the basic qualitative features can be obtained in the mimetic framework. Note that \cite{myrzakulov2016static} \cite{sebastiani2017mimetic} explained rotation curves through an additional potential term (as well as complex scalar field and/or space-like constraint), we seek to do so without any extra potential terms with a real mimetic field. A static space-time immediately implies that $E$ should be time-independent and a function of $r$ only, and that $\psi$ is a function of $r$ and possibly linear in $t$. We start by assuming it's independent of $t$. Then \eqref{eq:test2} implies:

$$
 \nabla \cdot (E\nabla\psi)=0$$ If we assume $E\nabla \psi$ vanishes at infinity then we obtain:

\begin{equation}\label{eq:test6} E\nabla\psi=\dfrac{\alpha \vec{r}}{r^3} \end{equation}where $\alpha$ is the free parameter of our theory. If we set it to zero this will immediately imply that $E$ is zero (otherwise \eqref{eq:test3} would be inconsistent), hence we assume it's not zero and proceed. In that case \eqref{eq:test3} reads:

 $$\frac{U}{c^2}+\dfrac{\psi'(r)^2}{2}=0$$Hence assuming $-U > 0$ we obtain:

$$\psi'(r)=\pm \frac{1}{c} \sqrt{-2U}$$Plugging that in \eqref{eq:test4} we obtain:

$$E=-\dfrac{\alpha c}{r^2\sqrt{-2U}}$$ where the sign ambiguity has been absorbed into $\alpha$ (the sign of $\alpha$ is chosen so that $-E$ comes out positive because this is the only physically relevant case). Plugging this back into \eqref{eq:test1} we arrive at the equation for the potential $U$:

\begin{equation}\label{eq:test7}\Delta U=4\pi G\rho+\dfrac{\alpha c}{r^2\sqrt{-2U}}\end{equation}

In general this has no closed form solution; hence we will obtain the solution numerically, but before we do so we can attempt to solve it approximately. To do this we first choose a characteristic length-scale $R$ of the solar system (say 1AU). Then we notice that $\alpha$ has dimensions of $v^2$; intuitively it is an energy/mass flux free parameter which represents the amount of mimetic contribution to the background. Hence by dimensional analysis we can compare it to a quantity made up of characteristic mass and lengths of the system: $\frac{GM}{R}$. If we assume that \begin{equation}\label{eq:test8}\alpha \ll \bigg(\frac{GM}{R}\bigg)^{\frac{3}{2}}\frac{1}{c}\end{equation} then it is apparent that we can solve the equation perturbatively. It will also turn out that this approximate solution will exhibit the basic qualitative features of the more precise numerical solution. Hence we solve the equations perturbatively. At zeroth order we have the well known solution:

$$U=-\dfrac{GM}{r}$$where 

$$M=\int \rho(x) d^3x $$ Then we plug this solution back in \eqref{eq:test7} to obtain:

$$\Delta U=4\pi G\rho+\dfrac{\alpha c}{r^2\sqrt{2\dfrac{GM}{r}}}$$Since we are interested in the behavior far away from the sources (where $\rho = 0$) we obtain:

$$\Delta U=\dfrac{\alpha c}{r^2\sqrt{2\dfrac{GM}{r}}}=\dfrac{\alpha c}{\sqrt{2GMr^3}}$$Since $U$ is only a function of $r$ we have:

$$\dfrac{1}{r^2}\dfrac{d}{dr}\bigg(r^2 \dfrac{dU}{dr} \bigg)=\dfrac{\alpha c}{\sqrt{2GMr^3}}$$ Integrating twice we finally obtain the full solution as

\begin{equation}\label{eq:test9}U= -GM/r +\alpha c \sqrt{\dfrac{8r}{9GM}}\end{equation}Clearly if $\alpha$ satisfies \eqref{eq:test8} then the second term is negligible on the solar system scale. In this way the mimetic fluid wouldn't affect the predictions on the solar system scale. However the hope is that it could play a role on galactic scales and explain the flat rotation curves. 
\par\vspace{15pt}
\noindent If we now apply the theory to the galactic scale and attempt (at least qualitatively) to reconstruct the dark matter potential, then $M$ would be the mass of the galaxy, and $R$ would be the characteristic length of the galaxy (say, the radius of visible matter). Then the second term must be of the same order as the first term (of course, then the perturbative analysis would break down but we'll do a numerical simulation shortly which will reveal the same qualitative features) at this radius $R$, from this one can determine $\alpha$: 

$$\alpha \approx \bigg(\frac{GM}{R}\bigg)^{\frac{3}{2}}\frac{1}{c}$$ Then from \eqref{eq:test9} we can see that the second term will dominate beyond the radius $R$. To recover this result in a more rigorous manner one needs to perform a numerical simulation. Recall, however, that if $\psi$ is time independent, then \eqref{eq:test3} implies that $U$ can't be positive. However one needs $U$ to have positive values to recover the correct galactic potential (with dark matter contribution). One can do this by assuming $\psi$ linearly depends on $t$. This would still lead to a static space-time, since only derivatives of $\psi$ appear in the equations. Hence if we assume that $\psi$ is of the form $\psi (t,r) = -kt + f(r)$ then \eqref{eq:test3} would imply $$f'(r)=\psi'(r)=\sqrt{\frac{-2U}{c^2}+2k}$$ then the modified Poisson equation to be solved is \begin{equation}\label{eq:test10}\Delta U=4\pi G\rho+\dfrac{\alpha }{r^2\sqrt{\frac{-2U}{c^2}+2k}}\end{equation} 
$k$ is dimensionless and must be of order $\frac{v^2}{c^2}$ to be consistent with the expansion. To get rid of $c$ let us redefine the variables $\alpha \to \alpha c$, so now $\alpha$ has dimensions of $v^3$ and $k \to kc^2$ so now $k$ has dimensions of speed squared and must be of order $v^2$, then the above equation becomes
\begin{equation}\label{eq:test11}\Delta U=4\pi G\rho+\dfrac{\alpha }{r^2\sqrt{-2U+2k}}\end{equation} We will solve this equation numerically. To do so we make the following further assumption: We assume that inside the radius $R$ only the first term contributes and so the second term is actually multiplied by a step function. Another way of saying this is as follows: We choose units in which $GM=1$ and $R=1$. Then we will simulate the equation 

\begin{equation}\label{eq:test12}\Delta U=\dfrac{\alpha }{r^2\sqrt{-2U+2k}}\end{equation} with the boundary condition $U(1)=-1$, $U'(1)=1$

\begin{figure}[H]
  \centering
  \includegraphics[scale=0.4]{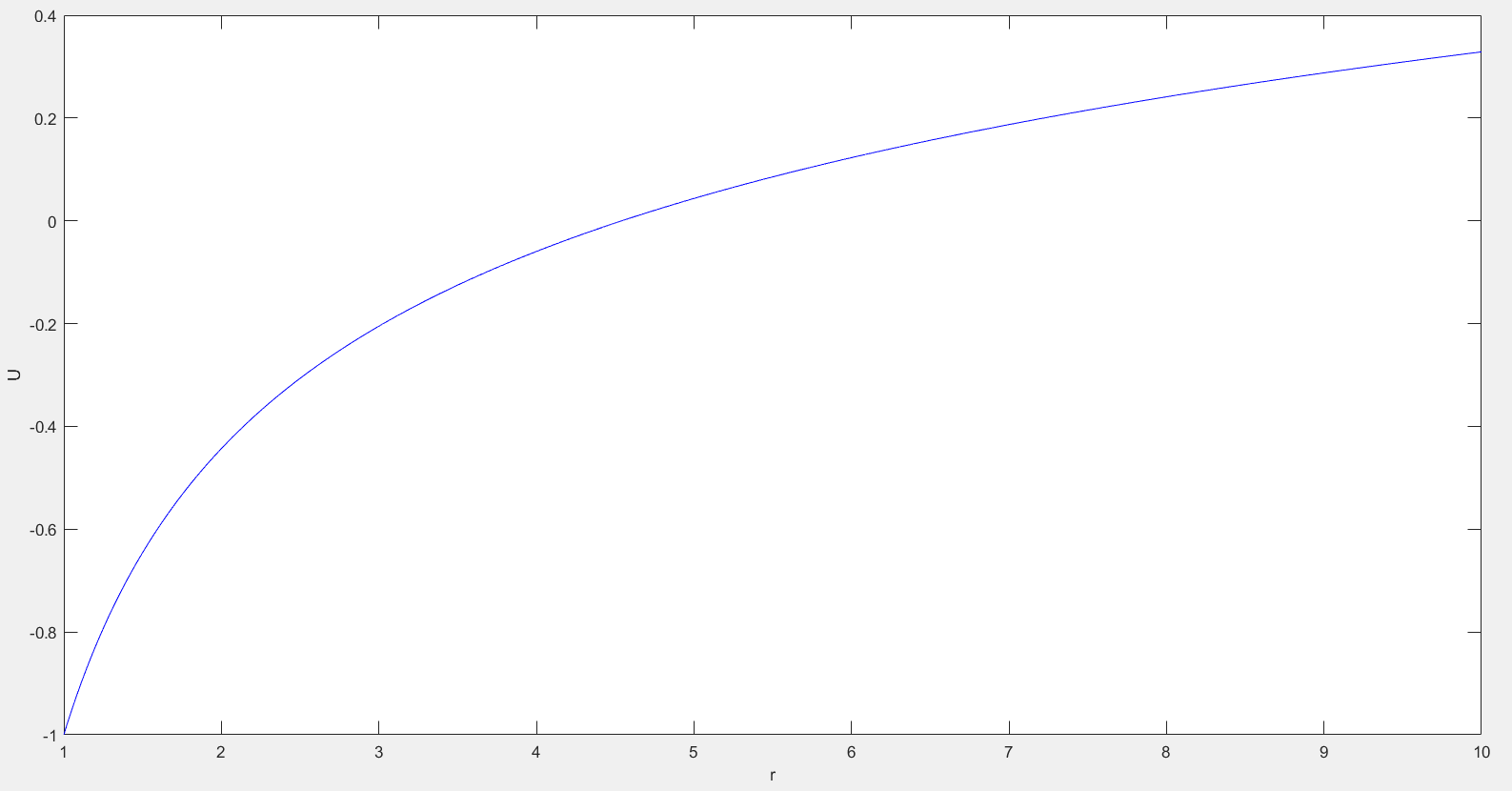}
  \caption{Numerical method 4th order Runge Kutta was used, with max step size 0.0001. The domain was taken to be [1,10] and the boundary conditions were $U(1)=-1$, $U'(1)=1$ in accordance with the standard Newtonian solution. $k=5$, $\alpha=1$.}   
  \label{fig:pic7}
\end{figure}

Note that when we choose $k=5$ what we actually mean is $k= 5\frac{GM}{R}$. We can also plot the energy density and compare it with a $\frac{1}{r^2}$ density.

\begin{figure}[H]
  \centering
  \includegraphics[scale=0.4]{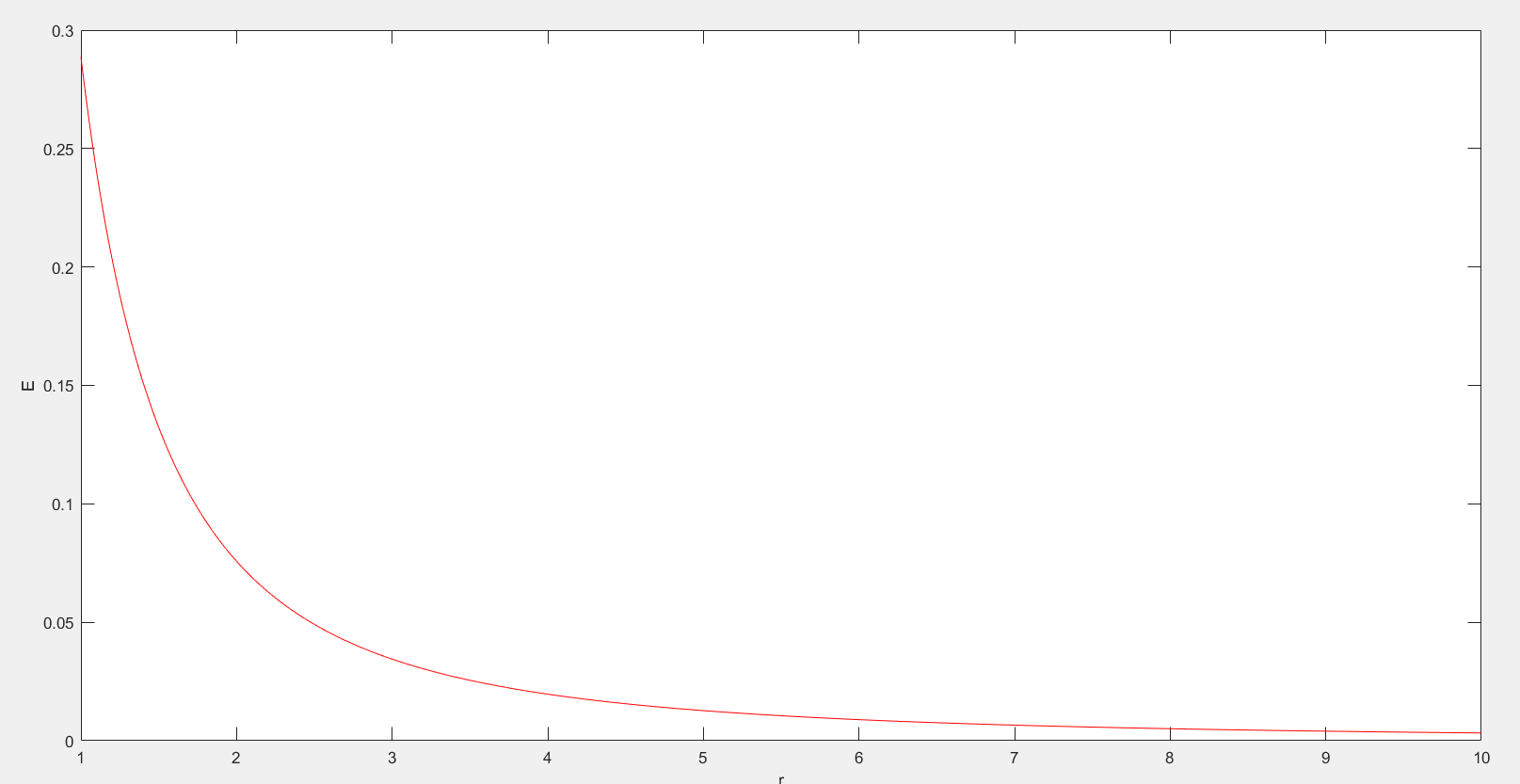}
  \caption{Plot of $E$ vs $r$ with domain size [1,10]. Same Boundary conditions as above.}   
  \label{fig:pic9}
\end{figure}

\begin{figure}[H]
  \centering
  \includegraphics[scale=0.4]{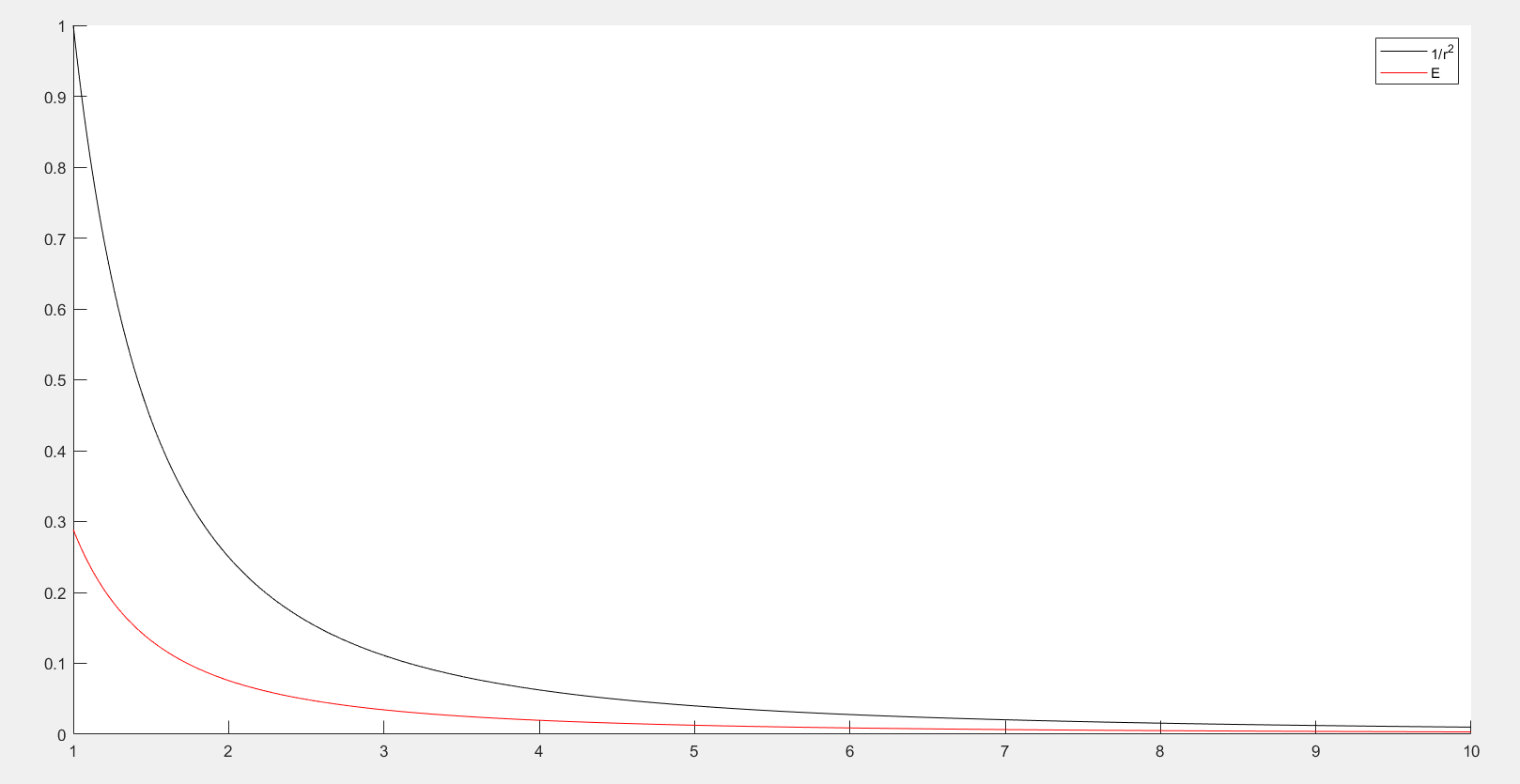}
  \caption{Comparison of $E$ with an inverse square density on a [1,10] domain.}   
  \label{fig:pic11}
\end{figure}

If we choose a bigger domain then the comparison would be as follows:

\begin{figure}[H]
  \centering
  \includegraphics[scale=0.4]{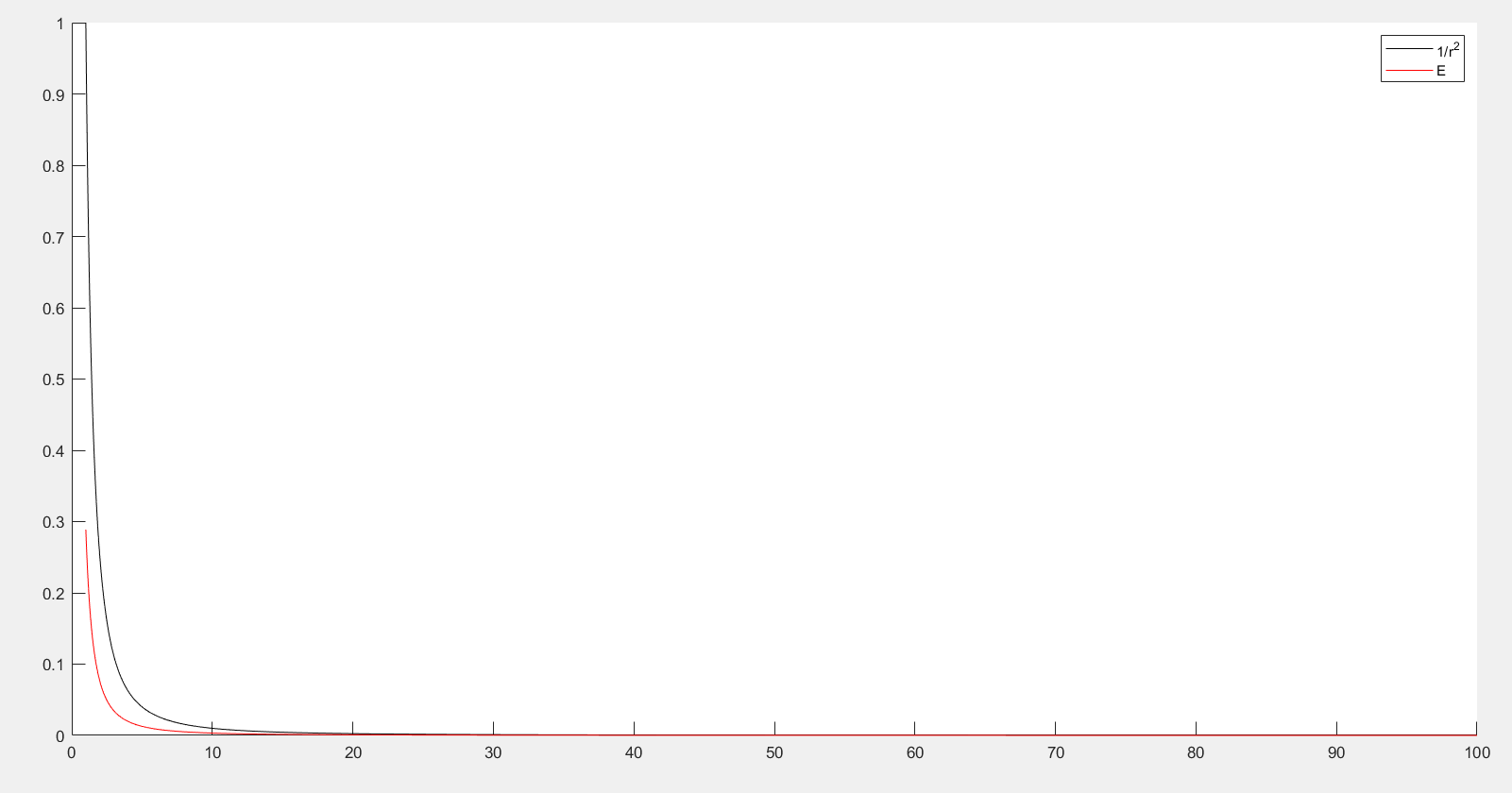}
  \caption{Comparison of $E$ with an inverse square density with [1,100] domain.}   
  \label{fig:pic13}
\end{figure}

As for the potential, choosing an even bigger domain reveals that this extra potential term has essentially a logarithmic dependence at large $r$:

\begin{figure}[H]
  \centering
  \includegraphics[scale=0.4]{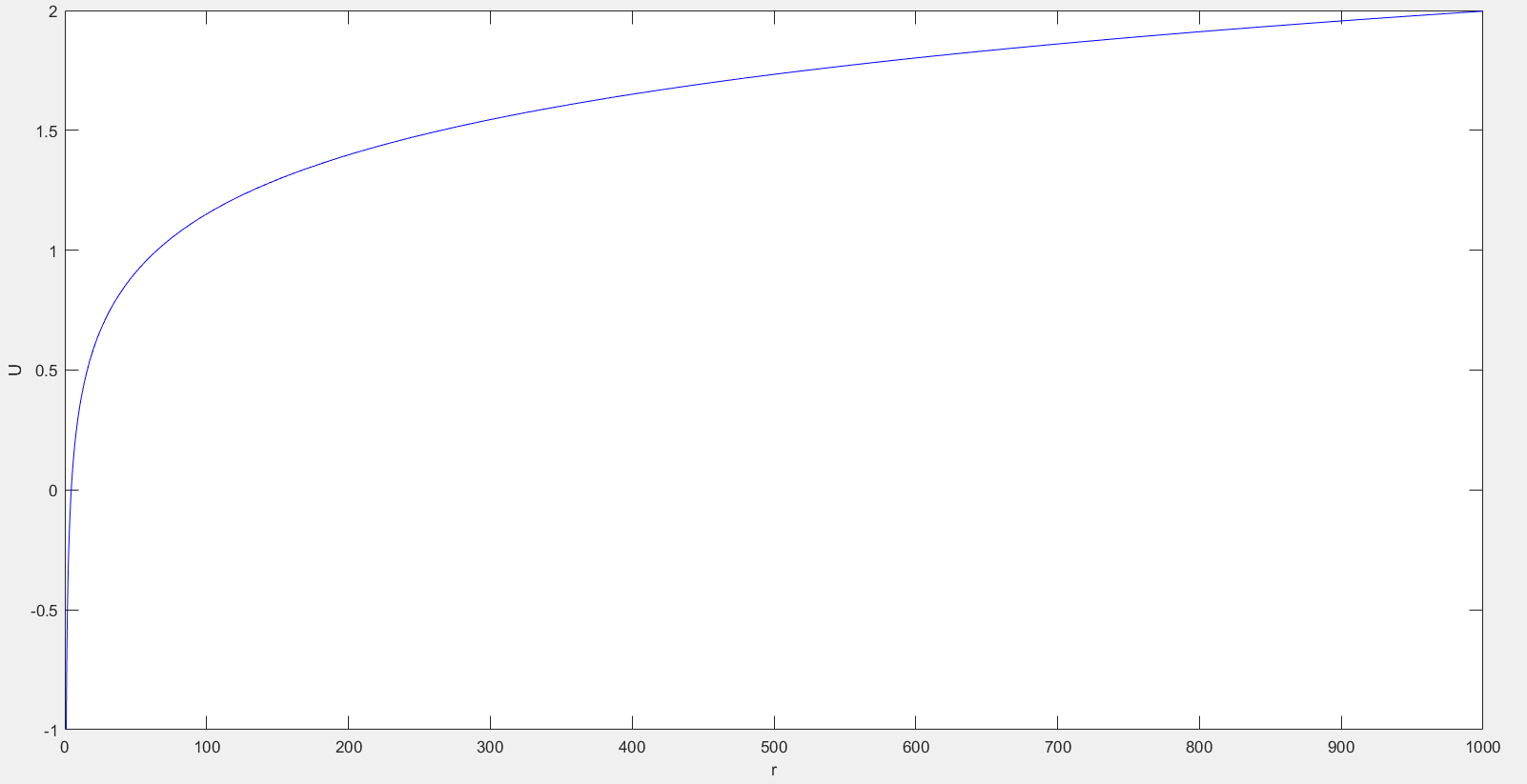}
  \caption{The domain was taken to be [1,1000] and the boundary conditions were $U(1)=-1$, $U'(1)=1$. k=5, $\alpha=1$.}   
  \label{fig:pic8}
\end{figure}

A direct comparison shows that it actually varies slower than a logarithm but faster than the square root of a logarithm. A logarithmic term is exactly what one needs to explain flat rotation curves. One can recover the basic qualitative features of the rotation curves by fixing particular boundary conditions:

\begin{figure}[H]
  \centering
  \includegraphics[scale=0.4]{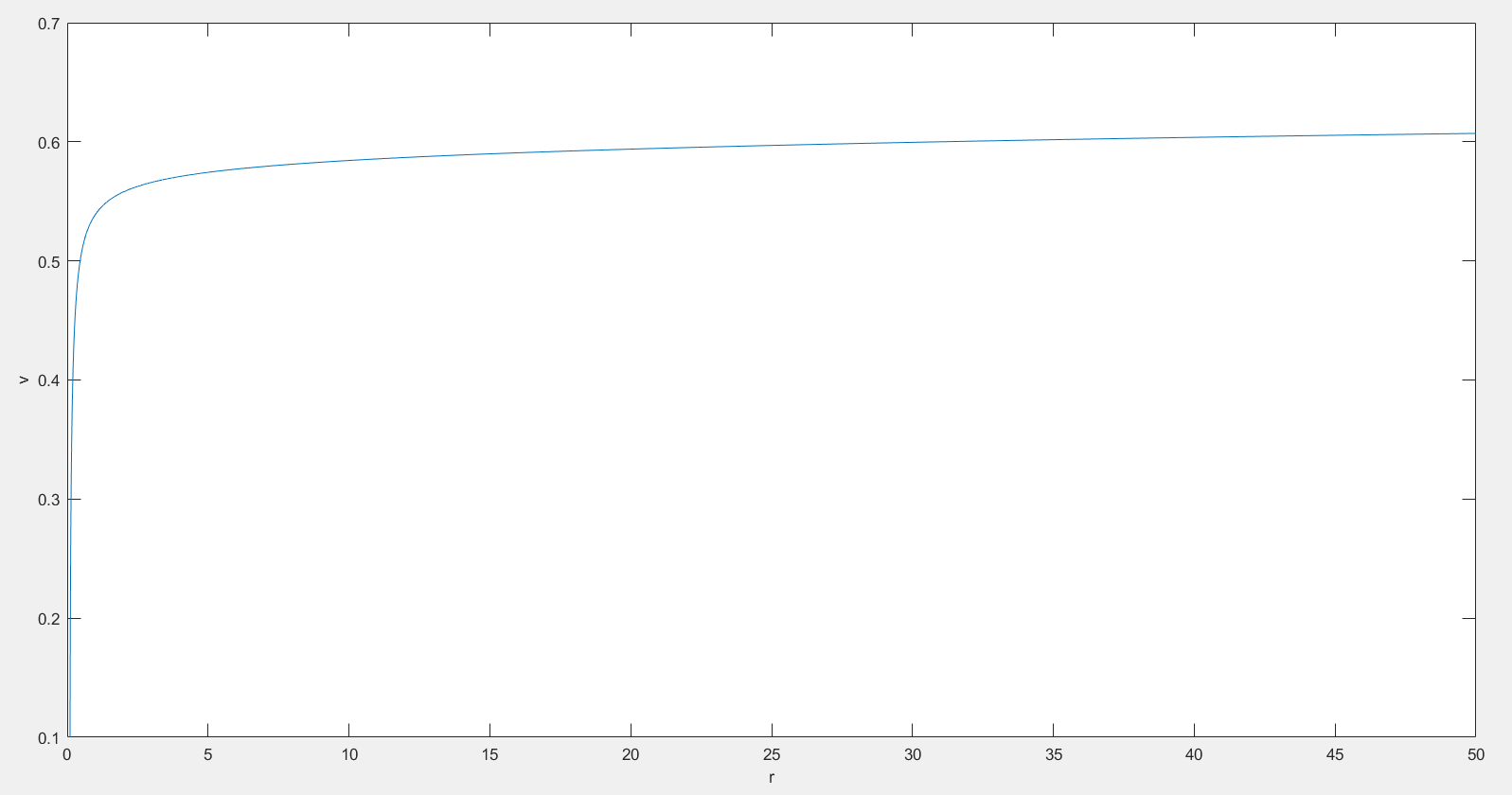}
  \caption{galaxy rotation curves. The domain was taken to be [0.1,50], the boundary conditions used are $U(0.1)=0.1$, $U'(0.1)=-0.1$.}   
  \label{fig:pic14}
\end{figure}

Fixing the boundary conditions amounts to fixing the amount of mimetic contributions inside the radius $r=0.1$. Here is the plot of the same boundary conditions with various values for $\alpha$.

\begin{figure}[H]
  \centering
  \includegraphics[scale=0.4]{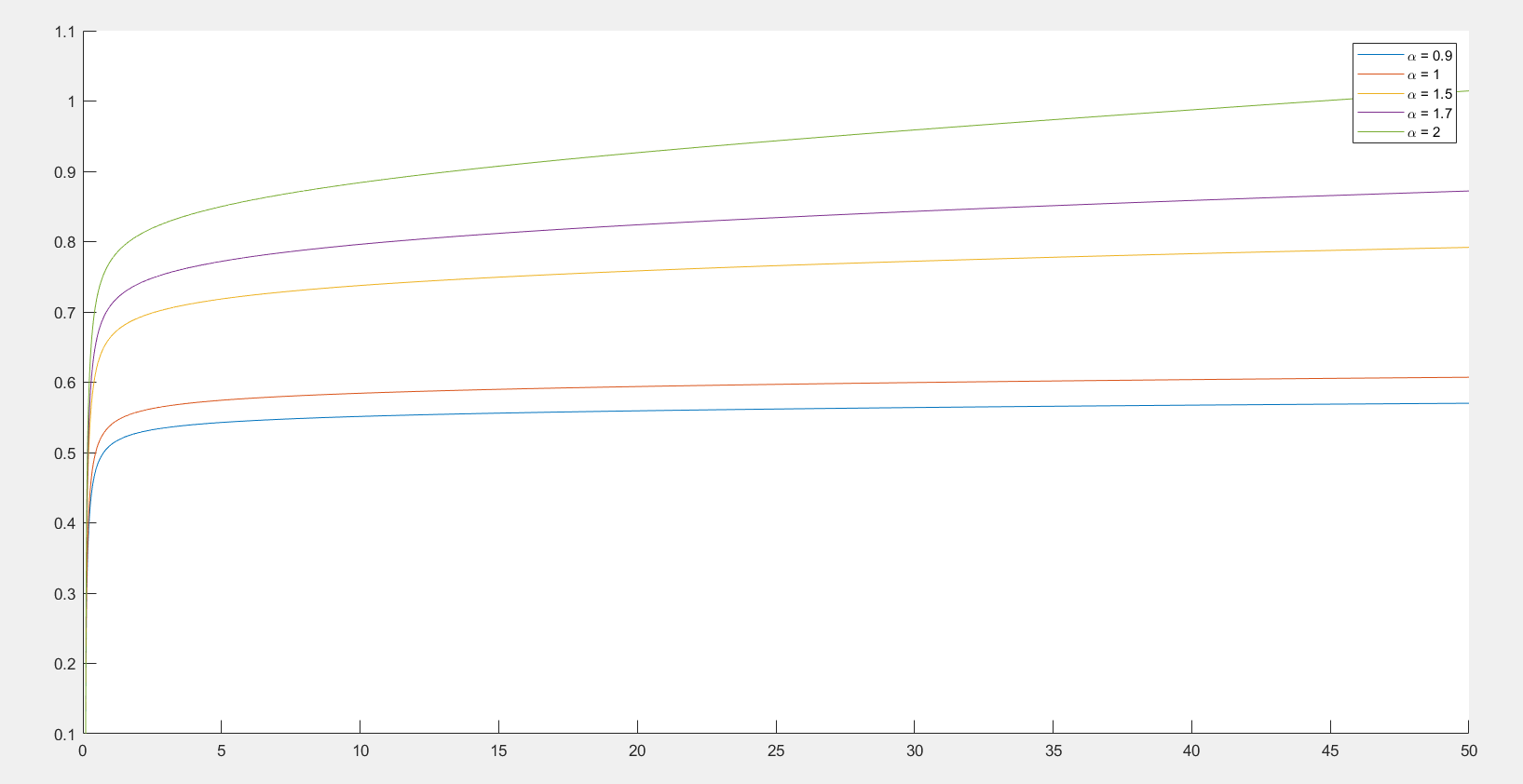}
  \caption{galaxy rotation curves. The domain was taken to be [0.1,50], the boundary conditions used are $U(0.1)=0.1$, $U'(0.1)=-0.1$.}   
  \label{fig:pic15}
\end{figure}

\section{Conclusion} We carried out the post-Newtonian expansion in Mimetic gravity. We were then able to derive the Newtonian limit of the theory. The mimetic contribution is characterized by an energy flux parameter; using this parameter we sought to mimic the effects of dark matter on astrophysical scales. We are able to recover the Newtonian solution as well as qualitatively explain flat rotation curves on galactic scales. To account precisely for the effects of dark matter one needs a logarithmic term in the potential; we show that even though such a term can't be obtained as an exact solution to the equations, by using numerical simulations one can obtain a quasi-logarithmic term in the potential. In this way, we are able to recover the flat rotation curves. We also prove a theorem that all static spherically symmetric solutions with non trivial mimetic contributions are not asymptotically flat.
\par\vspace{15pt}
\noindent To get an even better picture of where mimetic gravity stands regarding the galactic rotation curves, one needs to compare the numerical simulation curves to the most recent dark matter profiles/velocity curves coming from observational data. Note that one also has to solve the equations of mimetic gravity starting from inside the visible galactic matter (we only did so starting from the outside). 

\section{Acknowledgments}

I would like to thank Ali Chamseddine and Leonid Klushin for helpful discussions.

\bibliography{citations}
\bibliographystyle{plain} 

\end{document}